\documentclass[fleqn,10pt]{wlscirep}
\usepackage[utf8]{inputenc}
\usepackage[T1]{fontenc}
\usepackage{bm}
\usepackage{lineno}
\title{Quantum sensing 
for particle physics}

\author[1,2,*]{Steven D. Bass}
\author[3,+]{Michael Doser}
\affil[1]{Kitzb\"uhel Centre for Physics, Kitzb\"uhel, Austria.}
\affil[2]{
Jagiellonian University,
Marian Smoluchowski Institute of Physics and
Centre for Theranostics, 
Krak\'ow, Poland.}
\affil[3]{CERN, Experimental Physics Department, Geneva, Switzerland.}

\affil[*]{e-mail: Steven.Bass@cern.ch}
\affil[+]{e-mail: Michael.Doser@cern.ch}

\begin{abstract}
Quantum sensing is a 
rapidly growing approach to probe fundamental physics 
and explore new phase space for possible new physics
with precision and highly sensitive 
measurements in our quest to understand the deep structure of matter and its interactions.
This field uses properties of quantum mechanics in the detectors to go beyond traditional measurement techniques.
Key particle physics topics where quantum sensing 
can play a vital role include neutrino properties, 
tests of fundamental symmetries 
(Lorentz invariance and the equivalence principle
as well as 
searches for electric dipole moments and 
possible variations in fundamental constants), 
the search for dark matter  and testing 
ideas about  the nature of dark energy.
Interesting new sensor technologies include atom interferometry, optomechanical devices, 
and atomic and nuclear clocks including 
with entanglement.
This Perspective explores the opportunities for these technologies in future particle physics experiments,  
opening new windows 
on the  
structure of the Universe.
\end{abstract}

\begin{document}

\flushbottom
\maketitle

\thispagestyle{empty}

%

\section{Introduction}


Our present understanding of fundamental physics is built on the Standard Model, SM, of Particle Physics and General Relativity, GR, Einstein’s theory of gravitation.
Particle physics is described by the very successful SM
encompassing the theories of quantum electrodynamics, QED, electroweak interactions and quantum chromodynamics, QCD
\cite{Altarelli:2013tya}.
These interactions are governed by the gauge symmetries of U(1), chiral SU(2) acting on left handed quarks and leptons, and colour SU(3).
The masses of the 
W and Z gauge bosons as well as the quarks and charged leptons are determined by the 
Brout-Englert-Higgs, BEH, 
mechanism, with the discovery of the Higgs boson at CERN in 2012 completing the SM
\cite{Bass:2021acr}.
For gravitation
GR is working everywhere it has been tested, from 
laboratory experiments to astrophysics measurements, with recent highlights including gravitational waves and black hole imaging, plus gravitational lensing.

This is not the full story.
Open puzzles remain:
the origin of tiny neutrino masses 
(the lightest neutrino mass is expected to be about 10$^{-8}$ times the electron mass), 
the matter-antimatter asymmetry in the Universe, why the SM contains 3 families of quarks and leptons and, at the interface of particle physics and gravitation, the nature and properties of the mysterious dark energy, DE, and dark matter, DM, that comprise 68\% and 27\% of the energy budget of the Universe, with dark energy driving its accelerating expansion.

Might there be new interactions waiting to be discovered involving new particles 
to appear at higher energies or with tiny couplings, so called feebly interacting particles, that have not yet appeared with the present precision of our experiments?
The vacuum of the SM sits close to the border between stable and metastable if the SM is extrapolated up to Planck scale  
assuming no coupling to undiscovered new particles.
Might this be a clue? 
Might we find cracks in Standard Model symmetries with higher precision or energy, 
or perhaps with GR, 
e.g., 
as violations of the
equivalence principle, 
to give clues 
to a more fundamental theory?
Going further in our quest to understand the deep structure of matter and spacetime involves pushing 
the energy and precision of our experiments, and the interface with cosmology.
Many searches for new physics involve hunting for tiny cross sections, thus requiring new tools for precision measurements.

Advances with new 
quantum sensing technologies are driving 
the programme 
of precision measurements 
and are
also potentially relevant 
for high energy colliders and cosmology.
Quantum sensors are important in testing models of 
the dark sector
(dark energy and matter),
 spacetime symmetries, 
particle electric dipole moments and neutrino masses,
together
probing fundamental physics originating from a vast range of energy scales 
stretching from well below $10^{-14}$ eV 
(e.g. in the case of axion-like particle searches with spin-based sensors) 
to the Planck scale (e.g. in the case of optomechanical searches for ultra-heavy dark matter candidates). Given the relatively low cost of entry, the extraordinarily wide range of possible approaches, and the large gains in sensitivity, in explorable parameter space and in technological improvements that are within reasonable reach, it is not surprising that activities in these techniques are occurring around the globe in numerous universities and labs from eastern Asia, Australia, India, Europe to the Americas.

In this Perspective  we 
discuss 
a selection of key areas where 
quantum sensors can 
have a significant impact 
and recent 
detector developments and advances that have 
genuine potential to drive future experimental progress,
including new 
approaches where the use of
quantum entanglement in clocks and interferometers 
will make for even more accurate devices 
with new discovery potential.

We first give a brief summary of quantum sensing techniques in  particle physics and 
then highlight a selection of 
key physics observables
where quantum sensors have real promise for discovery science.
Specific detector issues for particle physicists are discussed. 
Finally, 
we 
conclude with challenges for the particle and quantum sensor  communities with
opportunities for new collaboration that could drive the search for new physics at the precision frontier.


\section{Overview of quantum sensor 
approaches}
For the purpose of this paper, we will consider as quantum sensors 
any device whose measurement capabilities are enabled by our ability to manipulate and / or read out its quantum states, 
such as, e.g., discrete vibrational, rotational or excitational energy levels of individual (natural or artificial) atoms, ions or molecules, but also individual or collective spin states, collective phononic excitation levels of ensembles, discrete changes in magnetic flux sensors and more generally any system whose discrete excitation values can be manipulated and / or probed with high precision. 
Furthermore, 
quantum mechanics is potentially used as a tool in detector technologies to go beyond 
different measurement 
limits~\cite{Clerk:2008tlb,10.1093/nsr/nwaa210},
e.g. 
the standard quantum limit
and shot noise limits 
of measurement uncertainty, 
with a view to approaching the fundamental Heisenberg limit imposed by the uncertainty relations
\cite{doi:10.1126/science.1104149}.
The already very high sensitivity of these systems to interactions can be enhanced even further when they are not considered in isolation but as an ensemble of entangled sub-systems or when they are prepared in such a manner that decoherence of the system's quantum states is minimized.

Quantum sensors fall naturally into two categories, those - at low energies - where the energy scale being probed is commensurate with those of the energy levels of the sensor itself, typically at or below the eV scale; and those where quantum systems form part of a larger system, in which their specifics enhance existing methods or enable novel types of detectors better suited to high energy particle physics (HEP), with energies above the keV scale.

Sensing technologies in the former realm and relevant to low energy particle physics include (but are not limited to):
atom interferometry~\cite{Overstreet:2020ftt}, 
quantum clocks ( atomic~\cite{Safronova2023}, ionic~\cite{Ion_clocks} and nuclear~\cite{Peik:2020cwm}),
optomechanical devices, 
spin-sensitive devices~\cite{PhysRevA.108.010101}, 
superconducting sensors,
superconducting RF cavities~\cite{arxiv:2203.12714}, 
and a wide range of technologies at the nanoscale~\cite{Quantum_nanotech}, including levitated nanoparticles.  
%
These technologies result in generally small-scale set-ups that focus on individual or small numbers of superconducting sensor elements or of trapped ions, atoms or molecules, on optomechanical sensors (which however can exceed the gram-scale), on atom magnetometers or sensors coupling to spin; 
they are however also relevant at the larger detector scales of HEP, such as for detectors relying on kinetic recoils in searches for light dark matter, and may even enhance the functionalities of particle detectors operating well beyond the level of sensitivity to individual quantum state changes~\cite{ECFA_roadmap}. 
A large number of reviews (inter alia Refs.~\cite{Safronova, HCI, Review_Snowmass_Maruyama_Kent}) has highlighted the rapidly expanding potential of these numerous accessible and complementary approaches in searching for new physics (e.g. in searching for ultralight scalar and vector dark matter\cite{Antypas:2022asj} or for ultra-light axions~\cite{heterodyne}), also in domains particularly relevant for HEP~\cite{Snowmass_chapter5, Cosmic_visions},
or in probing known physics with novel devices with greatly enhanced sensitivity, e.g. 
very low~\cite{10.1093/mnras/stab3418, martens2023lisamax} or high-frequency gravitational wave detectors\cite{HF_gravitational_waves}.

Before addressing higher energy applications of quantum systems, some of the central advantages that the above quantum systems bring to low energy particle physics deserve highlighting. In all of the above, state changes induced in the system through a (minute) external interaction lead to observable changes of the system's properties. The following examples are by no means exhaustive, but rather serve to illustrate the breadth of quantum systems that are of relevance to particle physics.

In the case of an atom interferometer~\cite{Bongs:2019}, a differential interaction between the two branches of an atom beam or of a trapped atom that is optically split into two states (a superposition of ground and excited state) in a first step and optically recombined at a later time will lead to phase shifts of the resulting interference fringes. The greater the separation in energy between the two possible states of the atom or paths the atom takes, the greater the sensitivity to e.g. gravitational disturbance during the time of state superposition. A terrestrial 
large baseline atom interferometer 
(vertical, such as AION~\cite{AION} or MAGIS~\cite{MAGIS} or horizontal, such as MIGA~\cite{MIGA}, 
all three under development or prototyping for larger baselines) 
will, once operational at the appropriate large scale, be  highly sensitive not only to low-frequency gravitational waves, but also to ultra-light dark matter~\cite{PhysRevD.107.055002} that would similarly couple to the atom during the time it is in a superposition of states.

In the case of continuously cooled superconducting sensors poised at the transition edge of switching to a normal-conducting state, the energy deposited by an interaction with a photon or the passage of a charged particle in the sensor can generate a brief, local switch to a normal conducting state, resulting in a brief voltage pulse before the sensor cools again to its initial superconducting state. 
Thanks to its very high sensitivity, even minute energy deposits, such as those from putative milli- or microcharged particles would be detectable.
Arrays of sensors are
mature technologies in standard use 
such as astrophysical imaging detectors for microwave photons. 
These include, 
e.g.,  transition edge sensors (TES), 
superconducting nanowire single photon detectors (SNSPD - in which the absorption of a photon briefly changes the resistivity of a circuit containing the nanowire) 
and 
kinetic inductance detectors, KID, where
absorbed photons produce excitations which change the resonance frequency and dissipation of a resonator incorporating a KID element. 
Element-specific resonance parameters allow frequency domain multiplexed (FDM) KIDs.
Compared to semiconductor pixel detectors, in which the the bandgap of ${\cal O}(1 {\rm eV} \sim 1000 {\rm nm})$ 
entails that low energy photons can only be detected with very poor energy resolution, the corresponding excitation process in superconductors (the breaking of Cooper pairs) requires ${\cal O} (10^{-3}$ eV), a significant boon for the spectroscopy of infrared  or microwave photons.

Optomechanical sensors~\cite{optomechanical_sensors} couple mechanical degrees of freedom to optical probes, as in e.g. optical cavities with deformable membranes~\cite{PhysRevLett.126.061301}, heterodyne cavities~\cite{heterodyne}, 
micro-mechanical oscillators~\cite{Riedinger} or optically levitated sensors~\cite{Geraci}.
The low coupling of these sensors to their environment (and thus their low decoherence) allows such sensors - once in their vibrational ground state - to be sensitive at the single phonon level. Their optical coupling can furthermore in some cases be exploited to achieve the necessary cooling of the sensor.
An interesting 
hybrid inertial sensing system that combines both optomechanical approaches and spin sensitivity is 
a spin-polarized nitrogen-vacancy diamond~\cite{NVdiamond} levitated in vacuum which can be briefly accelerated by a transient interaction coupling to its spin, leading to a transient minute shift in its continuously spectroscopically probed internal energy levels.

More generally, spin-based sensors~\cite{Geraci} probe a wide range of physics domains ranging from interactions between putative dark matter fields and the spin of standard model fermions, 
to novel spin-dependent interactions mediated by new light bosons to permanent electric dipole moments (EDMs, discussed below) that would violate discrete symmetries. Spin-based magnetometry that allows probing such interactions relies on numerous rapidly developing techniques, from nitrogen-vacancy diamonds to atomic Bose-Einstein condensate, BEC,  magnetometers to optical atomic magnetometers to single-domain ferromagnetic BEC magnetometers and most recently to levitated ferromagnetic torque sensors~\cite{suspended_ferromagnets}.

Finally, 
atomic, ionic, molecular or nuclear clocks measure time through the resonant frequency of the corresponding system-specific transitions.
Such transitions are highly sensitive probes of variations of fundamental constants (e.g. due to interactions with dark matter~\cite{Flambaum}), fundamental symmetries (e.g. Local Lorentz Invariance~\cite{LLI}) or searches for new particles (e.g. through isotope-shift atomic spectroscopy that could reveal new scalar or vector bosons~\cite{Frugiuele}) in addition to their potential for gravitational wave detection~\cite{Kolkowitz}. Identifying such perturbations in these systems involves comparisons of different types of clocks (or clocks , both locally and remotely, with different orientations, isotopic composition and reference systems.
%
Transient or oscillatory interactions affecting highly-precise atomic clocks (with frequency uncertainties at the level of $10^{-18}$ or less~\cite{PhysRevLett.126.011102}), will affect their clock frequency; interactions with dark matter would cause a persistent signal detectable by two clocks or clock/cavity systems, even in the same location. However, networks of clocks are not only highly sensitive to both such local (periodic or transient) changes  as to slow drifts of fundamental constants, but are all the more sensitive the larger the size of the network is, as spurious local causes are easily identified. 
More importantly, a distributed set of observations can allow identifying the temporal evolution and direction of a potential source behind common observations. 
Furthermore, as different atomic clocks (and the single candidate for a nuclear clock, $^{229m}$Th) 
have a wide range of different systematics and couplings, and thus sensitivities, to various putative beyond the Standard Model, BSM, physics, dedicated networks spanning 
continents or the globe, and composed of a large number of heterogeneous clocks would not only represent an ultimately sensitive system, but would also allow determining the directionality of e.g. domain wall-produced perturbations~\cite{Barontini_2022}.

The sensitivity of such systems can be further enhanced: a number of particularly promising developments build on quantum squeezing~\cite{squeezing} (where for a set of complementary variables, the component of most interest is measured to high precision at the expense of a large uncertainty in the other one); on entangling ensembles of quantum systems for distributed quantum sensing~\cite{distributedQS}; and on setting up arrays of such detectors, on increasing their accessibility through standardization, or on building nation- or continent-spanning networks of linked detectors~\cite{networks_atomic_clocks,networks_gravitational_waves, networks_magnetometers,network_dark_sector}.
It must be pointed out however that a number of other areas of possible improvement, such as increasing the number and coherence times, decreasing the temperature, and achieving higher control of possibly application-specific systems will likely lead to more straightforward initial gains.

In addition to low energy particle physics experiments, 
quantum sensing and the concomitant technologies also have implications for high energy physics detectors that will be expanded on in Section~\ref{HEPdetectors}. In these, the interaction of particles with multiple quantum systems has the potential to improve sensitivity, to provide a broader range of available technologies to measure a given quantity or to allow heretofore difficult or impossible measurements.

Beyond sensing, 
quantum computing ideas
are also discussed in particle physics, 
for example with processing 
huge data sets or in improving trigger or data reconstruction performance in future collider experiments like 
the high luminosity upgrade of the Large Hadron Collider 
\cite{Gray:2022fou}.
Quantum simulators \cite{Georgescu:2013oza}
are being developed as an alternative to usual lattice Monte-Carlo calculations for investigating non-perturbative properties of quantum field theories \cite{Banuls:2019bmf}.
Quantum sensors are also playing an essential role in driving particle physics applications in 
biomedical sciences 
with devices discussed 
in~\cite{9875702}
including the possible use of 
entanglement in next generation PET scanners\cite{Hiesmayr:2017xgx,Sharma:2022ehf}.

\section{
Key physics observables where quantum sensors can play a vital role}
%

We next list key observables 
where quantum sensing can play a vital role, with prime focus on neutrino properties,
fundamental symmetries including the equivalence principle, 
dark matter and dark energy.

\begin{itemize}
\item
{\bf {\it Neutrinos}}

In the minimal particle physics SM 
neutrinos come with 
left-handed chirality and with interact just through couplings to massive W and Z weak gauge bosons. 
In the minimal SM neutrinos come with zero mass. 
This assumption has proven too simple.
Neutrino oscillation experiments, 
where neutrinos created with a particular flavour (corresponding to electron, muon or tau) are later measured to have a different flavour, 
point to the existence of tiny neutrino masses.
Assuming three species of neutrinos,
the neutrino oscillation data 
constrains the largest 
mass squared difference
to be
$\approx 2 \times 10^{-3}$ eV$^2$ 
with the smaller one as 
$(7.53 \pm 0.18) \times 10^{-5}$ eV$^2$
\ 
\cite{BahaBalantekin:2018ppj}.
With these values the lightest neutrino mass is expected to be about 10$^{-8}$ times the value of the electron mass.
Mass measurements for individual neutrino species remain a major experimental challenge. 
Cosmology gives a bound on the mass sum with 
the Particle Data Group combined value ~\cite{Workman:2022ynf}, 
$
\sum m_{\nu} < 0.12 \ {\rm eV}$ 
in context of the 
usual $\Lambda$CDM model of cosmology with individual measurements typically bounded by about 0.5 eV.
Where do these masses come from?
A further 
open question is whether neutrinos are Dirac or Majorana fermions, with Majorana particles their own antiparticles. 
There are hints 
for charge-parity, CP, violation 
in the neutrino sector at the level of 1-2 $\sigma$ \cite{Workman:2022ynf}.
Being weakly interacting, precision measurements of neutrino properties are very challenging. 
If the neutrinos are Majorana 
with masses determined either through the 
so called see-saw mechanism or the mass dimension five 
Weinberg operator~\cite{Altarelli:2013tya}, 
then their masses are expected
to be 
$\sim \Lambda_{\rm ew}^2/M$
where $\Lambda_{\rm ew}$ is the
electroweak scale 246 GeV and $M$ is a large mass scale, typically about 10$^{16}$ GeV. 
Neutrino mass measurements
are thus exploring very high energy scales.
If neutrinos are Dirac particles, then one option is sterile right-handed neutrinos without direct coupling to Standard Model particles though here one needs some new mechanism to explain their tiny masses compared to charged leptons and quarks.

Majorana neutrinos 
have the signature of 
lepton number violating 
neutrinoless double $\beta-$decays involving the decay of two neutrons to two protons plus electrons with no neutrinos emitted, viz. 
$
(A,Z) \to (A, Z+2) + e^- + e^-$ 
where 
$A$ 
is the atomic mass number and 
$Z$ is the atomic number. %
Planned experiments 
using current technologies will reach the precision
$m_{\beta \beta} < 15$ meV limit 
with $m_{\beta \beta}$
the modulus of a linear combination of neutrino masses.
New ideas using superconducting sensors and multi-sensor phonon and photon imaging   
could reach a sensitivity 
$m_{\beta \beta} \sim 4-7$ meV, 
probing deeper into the range of likely neutrino masses
(including both normal and inverted mass orderings)
\cite{CUPID:2019imh,CUPID:2022wpt}.
Experiments aimed at a direct neutrino mass determination, independent of whether neutrinos are their antiparticles or not, 
study the capture of an inner shell electron with measurement of the released energy spectrum
$^{163}$Ho$+e^- \to ^{163}$Dy$+\nu_e+Q_{EC}$ 
with $Q_{EC}$ the electron capture energy for the neutrino 
\cite{Gastaldo:2013wha,Ullom:2022kai},
and the tritium process
T$_2 \to ^3$HeT$^+ e^- \bar{\nu}_e$ 
for the antineutrino with recent KATRIN results reported in~\cite{KATRIN:2021uub}, 
$m_\nu < 0.8$ eV.
The $^{163}$Ho experiments use cryogenic calorimeters whereas KATRIN uses a Penning trap in the set-up (though not quantum control).
Next generation experiments aim to reach precision 
$\delta m_{\nu} \sim 40$ meV 
(Project 8)
or better (CRESDA);
for a review see 
\cite{Canning:2022nye}. 
The ongoing HUNTER cold trapped Cs atom experiment is searching for possible sterile neutrinos with mass range 
5-100 keV~\cite{Martoff:2021vxp}.
Best exclusion limits on sterile neutrinos in the mass range 100-850 keV are given by the BeEST experiment in \cite{Friedrich:2020nze}.
Other experiments are suggested for capture of 
relic neutrinos from the early Universe to search for evidence of the cosmic neutrino 
background
~\cite{PTOLEMY:2019hkd}.
This would be a tritium observatory with 
signal for relic neutrino capture being a peak in the 
electron spectrum above the $\beta-$decay endpoint of 
the reaction 
${\rm \nu_e + ^3H \to ^3 He + e^-}$.

\item
{\bf \it Electric Dipole Moments and CP violation}

Understanding the 
matter-antimatter asymmetry in the Universe
requires some extra source of CP violation beyond the
quark mixing described 
by the Cabbibo-Kobayashi-Maskawa matrix in the SM.
Important observables are
particle
electric dipole moments, EDMs,
which probe
possible CP violation 
from any new physics induced anisotropy in the vacuum polarization of the measured particle.
The interaction with an electric field ${\bf E}$ is described by an interaction term 
$- {\bf d_E \cdot E}$
where ${\bf d_E}$ is the electric dipole
moment. 
Under time reversal 
${\bf E} \to {\bf E}$,
${\bf d_E}$ is proportional to the particle's 
spin vector which is odd under time reversal. Hence any finite electric dipole moment corresponds to a violation of 
time reversal invariance, T, symmetry and, 
through the fundamental symmetry of combined
charge, parity and time reversal invariance, 
CPT, to a violation of CP. 
Studying multiple systems is essential since the scale of possible new CP violation might be different for interactions coupling to leptons and quarks.

The present best precision measurement of the electron EDM~\cite{Roussy:2022cmp} is
$|d_e| < 4.1 \times 10^{-30} \ e{\rm cm}$
using HfF$^+$
(or 
$
    |d_e| < 2.1 \times 10^{-29} \ e{\rm cm}
$
combining HfF$^+$ and previous ThO 
experiments~\cite{Roussy:2022cmp,ACME:2018yjb})
constrains possible CP violation effects to a scale competitive with the LHC, 
and probes deep into the parameter space of new physics models at the TeV scale.
In EDM experiments
an applied electric field induces 
an energy shift for a given quantum state of the measured particle.
The system is spin polarized via optical pumping or some other hyper-polarization
technique such that the system is in a superposition of quantum states with opposite
EDM-induced energy shifts. 
A nonzero EDM then causes the polarized spins to
precess in the presence of an electric field.
The neutron EDM is measured using cold neutron beams or trapped ultra-cold neutrons.
The best present measurement
\cite{Pendlebury:2015lrz}
$
    |d_n| < 3.0 \times 10^{-26} \ e{\rm cm}
$
constrains any
strong CP violation induced by gluon topological effects,
with the so called QCD 
$\theta$ parameter
less than $10^{-10}$.
Next generation experiments 
with quantum technologies aim at improved precision of  
a factor of 10$^4$ with $d_e$ and factor of 100 with $d_n$
\cite{Alarcon:2022ero}.
Complementary to these measurements,
recent experiments addressing hadronic EDM's (e.g. of TlF~\cite{CENTREX}) or using short lived 
radioactive molecules such as RaF~\cite{GarciaRuiz:2019kwh} for eEDM searches or for precision discrete symmetry tests are driven particularly by the fact that nuclear deformations can provide additional dramatic enhancements in sensitivity~\cite{arXiv:2302.02165v1}.
Numerous similarly sensitive di-atomic (e.g. ThO, HfF$^+$, YbF, ...) or polyatomic molecules (such as YbOH) well suited to laser cooling~\cite{arXiv:2302.10161}, are under active investigation worldwide.

\item
{\bf \it The fine structure constant}

Precision measurements also allow one to  search 
for new physics by looking for evidence of new radiative corrections from interactions beyond the Standard Model. 
A key observable is the fine structure constant $\alpha$. 
One compares the value of 
$\alpha$ extracted from measurements of the electron's anomalous magnetic moment $a_e=(g-2)/2$ plus QED theory with measurements from atom interferometry experiments.
In QED 
the electron's 
$a_e$ 
is given by a 
perturbative 
expansion in $\alpha$ which is known to ${\cal O}(\alpha^5)$ precision 
plus tiny QCD and weak interaction corrections
\cite{Aoyama:2017uqe}.
It has most recently been measured~\cite{Fan:2022eto}  to a precision of 
$0.13 \times 10^{-12}$. 
Atom interferometry experiments measure 
heavy Cs or Rb atomic masses
through
recoil of a Cs or Rb atom in an atomic lattice. 
The fine structure constant 
$
\alpha^2
=
\frac{2 R_{\infty}}{c}
\frac{m_{\rm atom}}{m_e} \frac{h}{m_{\rm atom}} 
$
then involves this atomic mass measurement combined with 
other experimental quantities: 
the Rydberg constant $R_{\infty}$ 
and
the ratio of the atom to electron mass 
$m_{\rm atom}/m_e$  
($c$ is the 
speed of light and $h$ is Planck's constant).
Any radiative corrections from BSM physics will enter $a_e$ but not the atomic physics measurements of $\alpha$.
Substituting 
the measurements of $\alpha$ extracted 
from
Cs \cite{Parker:2018vye}
and Rb \cite{Morel:2020dww}
tabletop atom interferometric
experiments into the QED expansion for $a_e$ 
gives agreement 
to one part in $10^{12}$.
Below this level 
one finds tensions with the 
present Cs and Rb measurements at
2.5 and 1.5 standard deviations 
with opposite signs
calling for extra precision 
measurements.

Similarly, precision spectroscopy in trapped atoms, molecules, ions and in trapped highly charged ions (HCI - a recent extensive review can be found in~\cite{Kozlov:2018mbp}) 
with relative frequency uncertainties at the level of down to parts in 10$^{18}$ 
offers a natural test bed for variations of fundamental constants and also for hypothetical transient or slowly varying  cosmological fields or topological defects. Particularly in HCI's, optical transitions can be extremely narrow and - in comparison with current atomic clocks - are less subject to external perturbations. The wide range of these narrow transitions in HCI's provides for particularly high sensitivity to the value of the fine structure constant, but these systems are equally well suited for precision tests of QED in the nonperturbative regime and for dark matter searches, as highlighted in~\cite{Kozlov:2018mbp}.
%

\item
{\bf \it Spacetime symmetries}

Studies of spacetime symmetries probe the connection with gravitation and relativity.
Important probes are tests of the equivalence principle and Lorentz invariance.
The equivalence principle enters at three levels~\cite{Charlton:2020kie,Will:2014kxa}.
Weak 
or Galilean 
equivalence, the WEP, concerns the universality of free fall, that the inertial and gravitational masses should be equal.
The Einstein equivalence principle, EEP, involves the extension to GR type theories
and says that the outcome of 
non gravitational experiments in free fall should be independent of 
the velocity of the 
apparatus as well as its place in spacetime.
The EEP includes the following postulates (1) WEP, (2) Local Lorentz invariance (LLI): the outcome of
any local nongravitational experiment conducted in free fall is independent of the velocity and the orientation of the
apparatus and (3) Local position invariance (LPI).
Strong equivalence, the SEP, concerns the extension from test masses to self gravitating bodies.
With the SEP gravitation proceeds by minimal coupling via the spacetime connection with no direct matter to curvature coupling,  
there are no 5th forces
from extra scalar
gravitational interactions and 
the value of 
Newton's constant $G$ is independent of where it is  measured in spacetime.

The most accurate test of the WEP presently comes from 
the MICROSCOPE experiment in space~\cite{MICROSCOPE:2022doy} 
working at 
${\cal O}(10^{-15})$, 
improving on 
laboratory
torsion balance pendulum type experiments~\cite{Wagner:2012ui} with precision 
${\cal O}(10^{-13})$
and atom interferometer
 measurements at
$
{\cal O} (10^{-12})$
\cite{Asenbaum:2020era}.
A factor of 100 improvement on the MICROSCOPE result might be obtained using the 
proposed STE-QUEST experiment using quantum superpositions of cold atoms in space~\cite{Bassi:2022ste}.
WEP tests are now being extended also to antimatter systems, 
with a first direct measurement of the Earth's gravitational 
acceleration $g$ with antihydrogen to $\sim$ 30\% having been performed by ALPHA-g~\cite{ALPHA_g-gravity} and several complementary approaches (AEgIS~\cite{AEgIS}, GBAR\cite{GBAR} ) attempting  
to reach precisions of $\delta g/g \sim 1\%$ or better;  
further WEP tests with mixed matter-antimatter systems (positronium~\cite{Ps_inertial}, antiprotonic atoms~\cite{protonium}) are also under study. 
Indirect measurements with antiprotons~\cite{BASE} already define the sensitivity range for the direct measurements to exceed, with a claimed indirect limit of $\delta g/g < 3\%$ stemming from a comparison of the cyclotron frequencies of trapped (anti)protons (interpreted as clock frequencies in the Sun's gravitational potential).  
Standard and novel cooling techniques to far below sub-K temperatures will be needed to continue to drive advances beyond these first limits and will benefit additional systems which are slowly reaching the stage at which they can be considered for complementary antimatter tests of the WEP. Among these, antiprotonic atoms can additionally be used as test-beds for performing precision spectroscopy in search of BSM physics, and can also be a pathway for forming novel quantum systems, such as trapped fully stripped, unstable radioisotope HCI's or hydrogen-like ionic Rydberg systems~\cite{PhysRevC.107.034314}.

Atomic and nuclear clocks are used to make precision tests of the EEP.
With the EEP 
dimensionless quantities such as $\alpha$ and the ratio 
of the electron to proton mass,  
$\mu_{ep}$, 
should be velocity and spacetime position independent. 
Any time dependence in these quantities should show up in atomic and nuclear spectra, including clock transition frequencies.
If interaction with dark matter might change fundamental constants this would also change the rate at which a clock ticks.
Clock measurements are reviewed in~\cite{Safronova:2019lex}.
Prime atomic systems are Al, Sr and Yb clocks.
A recent precision measurement~\cite{Lange:2020cul}   
of possible time dependence 
in "slow drift" (LPI test)
measurements 
based on the E2/E3 transition of $^{171}$Yb$^+$ %
gives 
${\dot \alpha}/\alpha = (1.0 \pm 1.1) \times 10^{-18}$ {\rm year}$^{-1}$
and
${\dot \mu_{ep}}/\mu_{ep} = (8 \pm 36) \times 10^{-18}$  {\rm year}$^{-1}$.
Clock experiments test possible spacetime dependence
today.
They complement astrophysics 
constraints on possible variations between now and the early Universe  \cite{Hart:2017ndk,Ubachs:2017zmg}.
Clocks are also used to test
With Lorentz symmetry (LLI tests) there are precision tests with clock and trapped ion experiments.
Two single-ion Yb$^+$ clocks 
were observed to agree
at the $10^{-18}$
level 
over a half-year long comparison period, within the validity of their uncertainty budgets
\cite{Sanner:2018atx}.
The two clock ions were confined in separate ion
traps with quantization axes aligned along nonparallel directions. 
Hypothetical Lorentz symmetry violations would lead to sidereal modulations of the frequency offset,
which were observed
to be absent at the 
10$^{-21}$ level, 
putting strong constraints on Lorentz symmetry 
violating extensions of the SM.
Similar  limits on Lorentz violation also follow from a recent trapped ion experiments~\cite{Dreissen:2022}.)
Improved precision could come from networks of linked atomic clocks
as described in~\cite{Barontini:2021mvu}
or with experiments with atomic clocks in space \cite{Delva:2017znr,Schkolnik:2022utn}.
As discussed below, atomic clocks place strict constraints on ultralight dark matter candidates.

Beyond atomic clocks, 
possible 
nuclear clocks based on $^{229}$Th using quantum states inside the thorium nucleus are a topic of vigorous research
\cite{2021NatRP...3..238B}. 
The thorium nucleus is 1000 times smaller than the electron shell making it less susceptible to environmental fields.
The transition from the ground state to the long lived 
low-lying $^{229m}$Th
isomer has recently been measured\cite{Kraemer:2022gpi} to have an excitation energy of 8.338 (24) eV  
making it unique in that laser techniques ($\lambda = 149.7 \pm 3.1$ nm) can be applied to induce nuclear transitions.

\item 
{\bf 
\it Dark matter candidates}

Dark matter is suggested by the cosmic microwave background, 
gravitational lensing and galaxy properties   \cite{Bertone:2018krk}.
So far its presence is deduced only through gravitational interactions.
The substance of this DM is presently unknown with ideas including hypothetical 
new particles without direct coupling to photons and with 
individual candidate
masses ranging from $10^{-22}$ eV to $10^{15}$ GeV~\cite{Baudis:2018bvr}.
(There are also ideas based on primordial black holes formed in the very early Universe and modified gravity theory scenarios.)
There is a vigorous global program to look for possible DM particles, partly indicated in figure~\ref{fig:DM} which emphasizes the role
quantum sensors play in the search for DM candidates with 
tiny DM-SM couplings
over a wide range of possible DM particle masses 
from ultra-light (mass less than 1 eV) to
light (less than about a GeV) through to ultra-heavy (up to close to the Planck mass)  
\cite{Lehnert:2021gbj,Antypas:2022asj,Buchmueller:2022djy,Ebadi:2022axg,Windchime:2022whs,Essig:2022dfa,Adams:2022pbo,Jaeckel:2022kwg,Chou:2022luk}.
Measurement precision is vital given the (at best) very small coupling
between DM and SM particles.
The expected  improvement in sensitivities to dark matter candidates over a 
large possible mass range 
with new quantum sensing technologies is illustrated in Fig.~\ref{fig:DM} with much more detail provided in the references of the caption and recent reviews 
\cite{Chou:2022luk,Adams:2022pbo,Jaeckel:2022kwg}.

One popular candidate for dark matter involves possible 
new pseudoscalar particles called axions with masses typically less than the m eV range. 
Axion searches are a prime focus of investigation given their possible role in resolving both the strong CP puzzle (why gluon topological effects do not generate large CP violation in QCD) and dark matter~\cite{Adams:2022pbo}.
Here superconducting 
single-photon detector 
sensors and spin-based sensors play an essential role.
Recent quantum enhanced 
pseudoscalar 
axion(-like) particle 
searches are reported in 
\cite{HAYSTAC:2020kwv,Roussy:2020ily} 
including 
the use of squeezing techniques to reach a precision beyond the standard quantum limit in QCD axion searches~\cite{HAYSTAC:2020kwv}.
Dark matter axions would cause
precession of nuclear spins 
and can also be searched for by comparing simultaneous,
co-located interferometers using Sr atoms in quantum states with differing nuclear spins 
\cite{Graham:2017ivz}.

Ultralight dark matter can act as a background field
which induces oscillations in Standard Model parameters like $\alpha$ and $\mu_{ep}$~\cite{Arvanitaki:2014faa} 
probing possible linear and quadratic couplings to photons, quarks and gluons~\cite{Antypas:2022asj}.
Most recent optical clock constraints 
of photon couplings 
and nuclear radii oscillation constraints on gluonic couplings using Yb$^+$ experiments are reported in 
~\cite{Filzinger:2023zrs} and \cite{Banerjee:2023bjc} respectively for possible bosonic DM particle masses less than $10^{-17}$ eV.
The BACON collaboration, 
using a network of optical clocks based 
on Al, Sr and Yb 
have measured frequency ratios to ${\cal O}(10^{-17})$ precision~\cite{Beloy:2020tgz}, and give further strong constraints on the coupling of ultra-light DM to the photons.
Searches for ultra-light DM candidates could also be performed using the same apparatus used for
future cold atom interferometer measurements \cite{Badurina:2021rgt}
of gravitational waves
in the mid frequency range between the present LIGO/Virgo/KAGRA experiments and the future LISA mission.
These experiments, initially on Earth and later in space \cite{Alonso:2022oot}, could also probe possible modifications of GR due to any graviton mass or Lorentz violation as well as look for evidence of possible phase transitions in the early Universe, cosmic strings and primoridial black holes. Sensitivities are given in 
Fig. 3 of~\cite{Badurina:2021rgt}.

An interesting challenge with DM detection experiments comes when
scattering in crystals runs into the neutrino floor of background of neutrinos from the Sun.  
To get around this, new ideas include 
the development of directional detectors using solid-state sensing \cite{Ebadi:2022axg}.
Further improvements in light DM search experiments may follow from advances with creating quantum superposition states of bodies of increasing mass and complexity with macromolecular interferometry \cite{2022AVSQS...4b0502K}.

\item
{\bf \it Dark energy}

Within GR dark energy is described by a time independent cosmological constant
which 
acts like a vacuum energy density perceived by gravitation.
It is observationally  characterised by a scale of about 
0.002 eV \cite{Planck:2018vyg}, similar in size
to the range of values expected for light neutrino masses 
\cite{Altarelli:2004cp,Bass:2020egf}.
The cosmological constant would be the same everywhere in the Universe whereas dark matter clumps like normal matter.
In other ideas DE is described
by new light scalar fields
with one possibility being 
 so-called chameleon fields \cite{Burrage:2017qrf}.
New scalar components to gravity give rise 
to "5th forces" between normal matter that would violate the SEP.
Chameleon type models 
screen these forces in regions of high matter density like in the laboratory, enabling them to evade precision gravity tests.
Beyond pendulum experiments which tell us that Newton's law is working down to 
52 $\mu$m \cite{Lee:2020zjt},
5th force chameleon models are tested in experiments using atom interferometers, 
with light mass atoms essentially insensitive to the screening mechanism, 
strongly constraining these models 
\cite{Jaffe:2016fsh,Sabulsky:2018jma,doi:10.1126/science.aay6428}.
Ideas for future quantum 
optomechanical 
tests are discussed in~\cite{Qvarfort:2021zrl}.
Quantum bounce experiments to test
dark energy models using ultracold neutrons are discussed in~\cite{Sponar:2020gfr}.

\end{itemize}


\begin{figure*}[h!]
\centering
\includegraphics[width=0.85\textwidth]{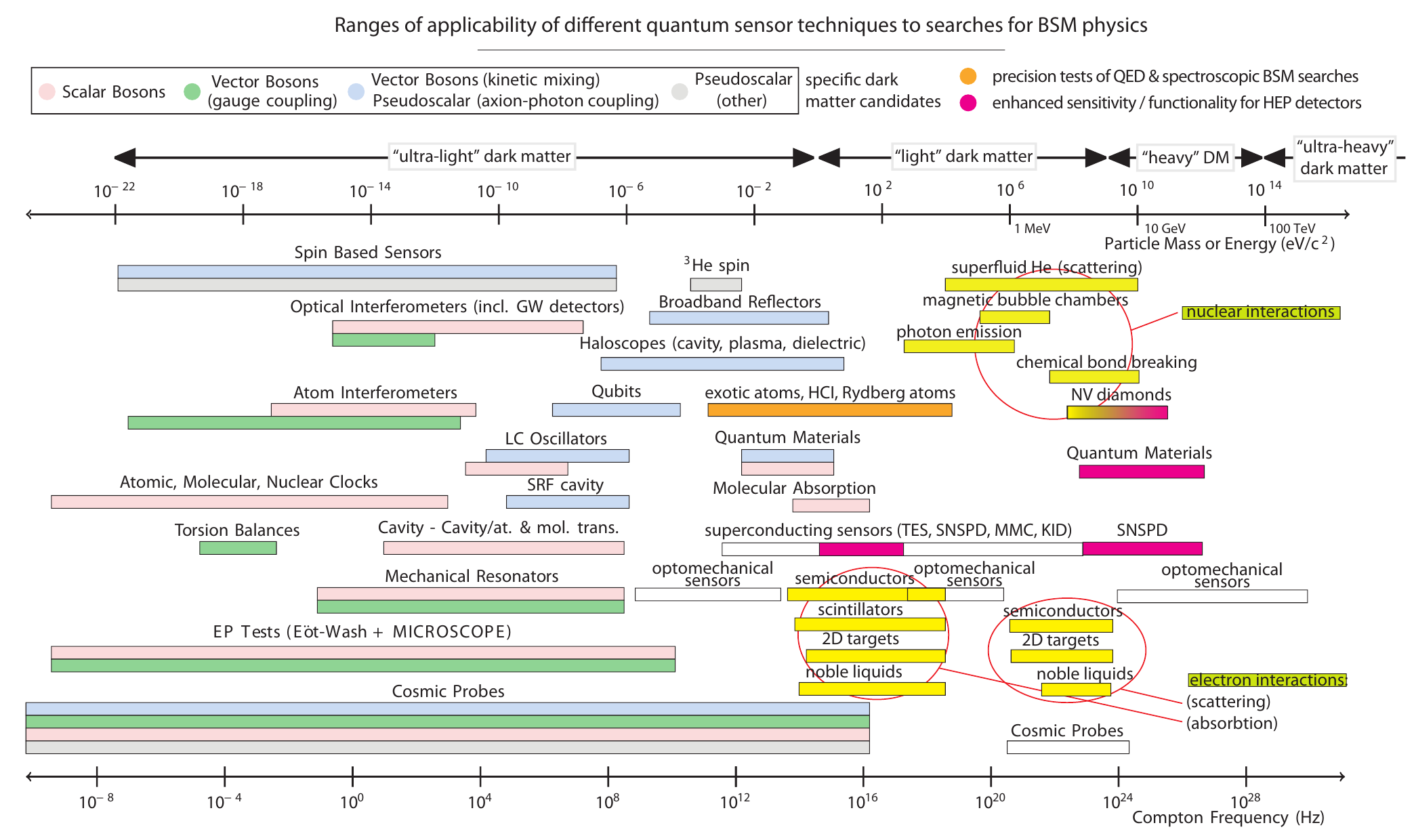}
\caption{
\label{fig:DM}
Ranges of sensitivity or applicability of a selection of different quantum sensors to searches for BSM physics.
While quantum sensors are mostly appropriate for searches for "ultra-light" and "light" dark matter, they also have potential applications as part of high energy physics detectors or to address "ultra-heavy" dark matter candidates. For several sensor types, typically probing symmetries or testing QED, the energies being probed are indicated rather than the masses of possible specific BSM candidates. Cosmic probes include observational measurements based on e.g. the CMB, cosmic ray upscattering, dwarf galaxies or the Lyman-$\alpha$ forest. 
Figure adapted from~\cite{Snowmass_chapter5}(fig. 5.12), with additional inputs from~\cite{Cosmic_visions}(fig. 4), and for 
NV diamonds~\cite{Marshall,Budnik},
optomechanical sensors at intermediate and high energies~\cite{arXiv:2306.09726v1,Moore_Geraci,Carney_QST2021}, 
cosmic probes at high energies~\cite{arXiv231104979} and others.
}
\end{figure*}


\section{Quantum sensing and high energy particle detectors}\label{HEPdetectors}

Contrary to the above examples from low energy particle physics experiments, where individual quantum sensors or quantum systems are well matched to the energy scale being probed, in high energy particle detectors, measurement of a particle's momentum or energy relies on repeated interactions between the particle to be measured and the sensitive material of a given detector.
 In these applications, it is often the bulk behavior of systems that can result from engineering at the atomic scale that can provide an extended functionality, can improve the sensitivity of existing devices or allow heretofore difficult or impossible measurements to be carried out, e.g. on the particle's helicity.
 
Attempts to improve the performance of calorimeters, charged particle trackers or different techniques that allow particle identification
by incorporating quantum dots\cite{Perovskytes2022} or two-dimensional molecular monolayers\cite{Orlandini2022} 
have 
only recently started, while devices capable of measurements of high energy photon polarization or particle helicity are only at the conceptual stage. The obvious potential for enhanced performance has thus led to formation of a number of new detector R\&D initiatives, among them the large-scale ECFA Detector R\&D roadmap\cite{ECFA_roadmap} or
new plans in connection with the Snowmass planning exercise in the United States
\cite{Buchmueller:2022djy,Cecil:2022fiv}, as well as initiatives such as the CERN Quantum Technology Initiative, QTI~\cite{DiMeglio:2021min}.
The 
just-beginning implementation of the ECFA detector R\&D for Quantum Sensing is expected to link numerous individual activities worldwide that focus on improved or novel detectors for particle physics, including those taking place in the contexts of numerous national (inter alia Quantum Vision in India, the Quantum Alliance in Germany, France Quantum, the UK National Quantum Technologies Programme or the National Quantum Initiatives in Japan or the USA) or supranational (Quantum Flagship in the EU) initiatives.

High-energy physics detectors are still undergoing vigorous research and development~\cite{ECFA_roadmap} in order to achieve the functionality required for future collider and fixed target experiments; exploration of the incorporation of quantum technologies could lead to improved functionalities beyond the present baseline. As part of this endeavor, developments stemming from the above initiatives can enhance the physics reach over existing techniques. 
Considerations on the potential of such approaches are recent: a small number of possible applications of low-dimensional systems (quantum dots, mono-atomic or multilayer structures), of metamaterials (e.g. crystal coatings), or of laser and microwave manipulation of large ensembles of atoms (e.g. in nitrogen-vacancy 
diamonds) have been proposed  
\cite{Doser:2022knm}, while an earlier overview of 
applications of quantum technologies to high energy physics~\cite{Ahmed:2018oog} highlighted their relevance
to highly sensitive searches for light dark matter candidates.

Specific quantum engineering-based detector developments could even lead to alternatives in the traditional design of complex multi-modal high-energy physics detectors.
To measure the identities and energies of both charged and neutral particles stemming from a common production point, these often consist of a central region of very low mass detectors (often silicon based) for minimally-perturbing charged particle trajectory determination surrounded by massive "calorimeters" (that measure, often via intermediate photon production, the ionization signals released as the particles scatter and lose their energy in them). Detecting any photons produced in the inner region would need photon detectors similar to those used in the calorimeter layer to be embedded in the inner detectors, resulting in massive degradation of the trajectory measurement capabilities, and is thus not considered. 
Given the transparency of e.g. silicon to infrared and microwave radiation, custom-built silicon detectors that combine charge collection with 
infrared transmission could however permit separating the light production region from the photodetection and amplification region, while "active scintillators" based on e.g. quantum dots, quantum wells or quantum well-dots, or devices combining collection of electrons liberated by ionizing radiation with quantum amplifiers (e.g. quantum cascade lasers) could open new possibilities in switching trajectory-measuring detector sensitivity at specific times ("priming") or building position-dependent-emission-frequency optical emitters that would be triggered by minimum ionizing particles (mip), thus resulting in "chromatic trackers". 
Finally, $ps$ timing jitter detectors, such as SNSPD's, could - in the form of multilayered charged particle tracking detectors - help resolve high pile-up rate situations such as those encountered in diffractive scattering events at high luminosity colliders such as the LHC or the EIC. All such novel devices would extend the range of observables for charged particles without negatively affecting overall performance. Of particular promise are thin-film photon and charged particle detecting devices with higher T$_C$ (up to 20K in the case of MgB$_2$~\cite{arxiv:2308.15228}), allowing to potentially avoid the need for dedicated and bulky dilution refrigerators.

In addition to the general-purpose possibilities mentioned above and discussed in detail in~\cite{Doser:2022knm}, a number of more specialized detector applications and improvements based on quantum systems, and which permit exploration of new phase space, can also be contemplated or are under active development. 
In these cases, the goal is to match the energy scale of the quantum detector with that of particle interactions within it 
(as is the case for hypothetical feebly interacting particles that would lose few tenths of eV, or x-rays depositing few keV), or to match the physical dimensions of the quantum system 
(typically at the nm scale) to those of the process under investigation, e.g. in the case of very short-lived particles. These two categories are well suited to incorporation in 
beam dump experiments (i.e. detectors placed behind massive particle absorbers) that investigate the wide range of secondary particles produced by a high energy particle beam interacting with a target upstream of the absorber) to search for millicharged 
dark matter particles~\cite{nanocharged}, complementary to similar searches with ion traps~\cite{nanocharged_ion_traps} or to search for very short lived particles.
Similarly, 
X-ray spectroscopy based on 
superconducting 
calorimeters~\cite{quantum_calorimeters} matched to the ${\cal O}$(10 keV) nuclear transition energies achieves eV energy resolution, an improvement of over one order of magnitude over traditional Ge-based x-ray detectors.
The use of stacks or layers of superconducting Josephson junctions or TES arrays in beam dump experiments could allow conceiving of a tracking detector for massive very minimally ionizing or lightly ionizing (milli-charged) particles. With a possible sensitivity of down to 12 meV/100 $\mu$m, such a device would be able to detect charges down to e/10$^6$ (the energy deposit of a mip being 20 keV/100 $\mu$m in e.g. silicon) and would provide for a beam dump experiment alternative to ongoing searches for similar heavy (mass > 1 MeV) particles, with a present sensitivity to lightly ionizing particles with charges as low as e/1000~\cite{Majorana:2018gib,Davidson:2000,Chang:2018}.

Entanglement or quantum correlations have only started being used in low energy particle physics, and have barely  
played a role in 
high-energy physics (with few exceptions, such as in the use of entangled kaons when constraining violations of the CPT symmetry~\cite{AMBROSINO2006315} by the DA$\Phi$NE experiment or in the observation - but not yet exploitation - of spin entanglement of top-quark pairs by the ATLAS experiment~\cite{arXiv:2311.07288}), 
although recent experience with atomic interferometers~\cite{Anders_PRL127}, atomic clocks~\cite{Nichol:2021lno}, trapped ions~\cite{Yb_entangled} or atoms in cavities~\cite{Greve:2021wil} indicates that limits obtained in entangled systems already exceed those of single quantum sensors by factors of two or more. The challenges then are, on one hand, to bring entanglement to bear on wide-spread networks of individual quantum sensors, and on the other, to identify situations in which entanglement can play a role in the context of high energy physics detectors.

\section{Conclusions and Outlook - challenges for the particle physics and quantum sensing communities}

The very rapid pace of advances in the field of Quantum Sensing has opened a large number of opportunities for exploration of new regions of phase space in particle physics, for exploiting highly sensitive probes of fundamental symmetries, and for developing novel approaches both in low and high energy particle physics. This pace continues to accelerate, with networking of individual quantum sensors, or even long-range entanglement, promising even higher sensitivities in the coming years.

In particular, quantum sensors have been and will continue to play a central role in tackling 
some of the most salient physics opportunities and challenges in the coming decade(s): neutrino masses, DM searches, search for BSM couplings, precision tests of fundamental symmetries and many more.
The feasibility of the concepts enumerated in the previous sections and that touch upon both low and high energy particle physics relies on active, targeted and shared R\&D on quantum sensors and on their ecosystem. Given that the challenges often lie between disciplines or require pooling of resources, collaborative approaches at a global scale appear to be ideally suited to bring about the necessary technological advances or improved infrastructure required to match the visions indicated in~\cite{ECFA_roadmap}. Further work that builds on existing expertise and ongoing developments would have major impact on, inter alia, the following:

\begin{itemize}
    \item
    {\underline{Networks, signal and clock distribution}: both time-stamping as well as distribution of highly accurate clock signals beyond national infrastructure or across continents opens the possibility for global networks of a wide range of experiments with different sensitivities to putative novel physics;}
    \item %
    {\underline{Enhanced cryogenic systems}: numerous engineering, material science and physics challenges remain that limit wider access to cryogenic systems, among them standardized and integrated electronics that operate at liquid He temperatures, superconducting devices that explore the trade-off between ultimate sensitivity and operation at higher (liquid helium or even liquid nitrogen?) temperatures, standardized approaches towards large-scale integrated systems or reduced access costs to sub-100 mK operation;}
    \item %
    {\underline{Exotic quantum systems}: novel atomic, (poly-atomic or radioactive) molecular and ionic systems, possibly in highly excited states, complex ensembles of heterogeneous trapped elementary particles, hybrid matter-antimatter bound systems; all these offer a rich palette for a wide range of precision spectroscopic probes of the standard model and of our understanding;}
    \item %
    {\underline{Theoretical developments}: both technical developments regarding predictions for systems amenable to precision measurements as well as identification of under- or unexplored phase space and the optimal systems to investigate those are crucial in optimizing resources;}
    \item %
    {\underline{"Bulkification \& large assemblages}"}: development of large--mass devices (such as macroscopic structures built of spin-polarized systems, e.g. as polarized scattering planes based on nitrogen-vacancy 
    diamond nanocrystals), or incorporation of quantum systems into existing bulk devices (e.g. tailor-built heterostructures or quantum dots for enhanced or novel functionality high energy physics  detectors); in parallel, very large component--number assemblies of individual quantum systems that maintain the individual component's specific properties; %
    \item %
    {\underline{Capability growth}: 
    improved opportunities for shared infrastructure (fabrication, tests, standardization), for simplified exchanges between the numerous developments in nano-engineered systems with novel quantum properties on the one hand, and particle physicists with unconventional requirements on the other hand, and coordinated educational initiatives will be crucial. Such efforts, such as a global quantum sensor R \& D collaboration (DRD5) under the aegis of ECFA~\cite{ECFA_roadmap}, are currently under preparation.}
\end{itemize}
Many of the relevant developments are already being tackled in the areas of astrophysical detectors, of material sciences, of cryogenic electronics, of quantum optics, and of high precision clocks. 
In some cases, %
it could well be that the collaborative approaches that have been so successful in high energy physics may bring about advances in the development and standardization of the tools and techniques needed for the exploration of the new particle physics parameters spaces touched upon in this Perspective. With a view more towards high energies, open questions remain surrounding which additional high energy physics detector approaches beyond those already highlighted could benefit from targeted R\&D, and perhaps more fundamentally, what role entanglement could play for high energy physics detectors? 
Given the very rapid growth of quantum technologies and their enthusiastic uptake and development for fundamental physics, it is clear that these will play a dominant role within the expanding field of particle physics in the coming decades.


\section*{Acknowledgements}

This article developed from discussions at the 
Humboldt Kolleg conference, 
{\it
Clues to a mysterious Universe - exploring the interface of particle, gravity and quantum physics}, 
Kitzb\"uhel June 26-July 01 2022.
The work was supported in part by 
the SciMat and qLife Priority Research Area
budget under the program Excellence Initiative - Research University at Jagiellonian University.

\bibliography{sample}

%

%


\end{document}